%% file: bouajjani.tex
\documentclass[envcountsame,envcountsect]{llncs}

\usepackage{latexsym,amsmath,amssymb} 
\usepackage[dvips]{graphicx}
\usepackage[dvips]{color}

\newcommand{\maps}[1]{\ {\overset{#1}{\longmapsto}}\ }
\newcommand{\hra}[2]{\ {\overset{#2}{\underset{#1}\hookrightarrow}}\ }
\newcommand{\lra}[2]{\ {\overset{#2}{\underset{#1}\longrightarrow}}\ }
\newcommand{\lrb}[1]{\ {\overset{#1}{\longrightarrow}}\ }
\newcommand{\leads}[2]{\ {\overset{#2}{\underset{#1}\leadsto}}\ }

\renewcommand{\top}{\mathit{top}} \newcommand{\pop}{\mathit{pop}}
\newcommand{\push}{\mathit{push}} \newcommand{\pre}{\mathit{pre}}
\newcommand{\post}{\mathit{post}}

\newcommand{\A}{\mathcal{A}} \newcommand{\B}{\mathcal{B}}
\newcommand{\C}{\mathcal{C}} \newcommand{\D}{\mathcal{D}}
\renewcommand{\H}{\mathcal{H}} \renewcommand{\S}{\mathcal{S}}
\renewcommand{\flat}[1]{#1\!\!\downarrow}
\newcommand{\sflat}[1]{#1\!\downarrow}

\author{Ahmed Bouajjani and Antoine Meyer}

\institute{{\sc Liafa}, Univ. of Paris 7, Case 7014, 2 place Jussieu
  75251, Paris Cedex 5, France
  \\
  \email{$\{$abou,ameyer$\}$@liafa.jussieu.fr}}

\title{Symbolic Reachability Analysis of Higher-Order Context-Free
  Processes}

\begin{document}

\maketitle

\begin{abstract}
  We consider the problem of symbolic reachability analysis of
  higher-order context-free processes.  These models are
  generalizations of the context-free processes (also called BPA
  processes) where each process manipulates a data structure which can
  be seen as a nested stack of stacks. Our main result is that, for
  any higher-order context-free process, the set of all predecessors
  of a given regular set of configurations is regular and effectively
  constructible. This result generalizes the analogous result which is
  known for level 1 context-free processes. We show that this result
  holds also in the case of backward reachability analysis under a
  regular constraint on configurations. As a corollary, we obtain a
  symbolic model checking algorithm for the temporal logic
  $\mathsf{E(U,X)}$ with regular atomic predicates, i.e., the fragment
  of CTL restricted to the $\mathsf{EU}$ and $\mathsf{EX}$ modalities.
\end{abstract}

\section{Introduction}
\label{sect-intro}

Pushdown systems and their related decision and algorithmic analysis
problems (reachability analysis, model checking, games solving and
control synthesis, etc) have been widely investigated in the last few
years \cite{caucal92,bcs96,walukiewicz96,bem97,ehrs00,cachat02,aem04}.
This recent intensive research effort is mainly motivated by the fact
that pushdown systems are quite natural models for sequential programs
with recursive procedure calls (see e.g., \cite{ek99,esparza02}), and
therefore they are particularly relevant for software verification and
design.

Higher-order pushdown systems \cite{engelfriet83} (HPDS) are
generalizations of these models in which the elements appearing in a
pushdown stack are no longer single letters but stacks themselves. We
call this kind of nested stack structures \emph{higher-order stores}.
Stores of level $1$ are sequences of symbols in some finite alphabet
(those are standard pushdown stacks), and stores of level $n+1$ are
sequences of stores of level $n$, for any $n > 1$. The operations
allowed on these structures are (1) the usual $\push$ and $\pop$
operations on the top-most level 1 store, (2) higher-order $\push$ and
$\pop$ operations allowing to \emph{duplicate} or \emph{erase} the
top-most level $k$ store of any given level $k \leq n$.

This general model is quite powerful and has nice structural
characterizations \cite{caucal02,cw03}. It has been in particular
proved in \cite{knu97} that HPDS are equivalent to (safe) higher-order
recursive program schemes.  Interestingly, it has also been proved
that the monadic second-order theory of an infinite tree generated by
a HPDS is decidable \cite{knu97,caucal92}, which generalizes the
analogous result for pushdown systems proved by Muller and Schupp
\cite{ms85}.  Also, it has been proved that parity games can be solved
for HPDS \cite{cachat03}, which generalizes the result of Walukiewicz
for pushdown systems \cite{walukiewicz96}.  These results actually
show that model checking is decidable for HPDS. However, they only
allow to check that a property holds in a {\em single} initial
configuration and they do not provide a procedure for computing a
representation of the set of configurations which satisfy some given
property (the satisfiability set of the property).

The basic step toward defining an algorithm which effectively computes
the satisfiability sets of properties is to provide a procedure for
computing the set of backward reachable configurations from a given
set of configurations, i.e. their set of predecessors. In fact, the
computation of forward- or backward-reachable sets is a fundamental
problem in program analysis and in verification.

Since HPDS are infinite-state systems, to solve this problem we need
to consider {\em symbolic representation structures} which (1) provide
finite representations of potentially infinite sets of configurations,
and (2) enjoy closure properties and decidability properties which are
necessary for their use in verification. Minimal requirements in this
regard are closure under union and intersection, and decidability of
the emptiness and inclusion problems.

A natural class of symbolic representations for infinite-state systems
is the class of finite-state automata. Recently, many works (including
several papers on the so-called \emph{regular model-checking}) have
shown that finite-state automata are suitably generic representation
structures, which allow to uniformly handle a wide variety of systems
including pushdown systems, FIFO-channel systems, parameterized
networks of processes, counter systems, etc.
\cite{bem97,bgww97,kmmps97,abj98,wb98,bjnt00,bouajjani01,hjjkprs95}.

In particular, for the class of pushdown systems, automata-based
symbolic reachability analysis techniques have been developed and
successfully applied in the context of program analysis
\cite{bem97,ehrs00,schwoon02}. Our aim in this paper is to extend this
approach to a subclass of HPDS called \emph{higher-order context-free
  processes} (HCFP for short). This class corresponds to the higher
order extension of the well-known context-free processes (also called
BPA processes). HCFP can actually be seen as HPDS with a single
control state, similarly to level 1 CFP which are equivalent to level
1 PDS with a single control state. The contributions of our paper can
be summarized as follows.

First, we observe that, due to the duplication operation, the set of
immediate successors (i.e. the $\post$ image) of a given regular set
of configurations is in general {\em not} regular, but it is always a
context-sensitive set.

Then, we prove that, and this is our main result, for every HCFP of
any level, the set of all predecessors (i.e. the $\pre^*$ image) of
any given regular set of configurations is a regular set and
effectively constructible.  As a corollary of this result, we obtain a
symbolic model checking algorithm (an algorithm which computes the set
of all configurations satisfying a formula) for the temporal logic
$\mathsf{E(F,X)}$ with regular atomic predicates, i.e., the fragment
of CTL with the modalities $\mathsf{EF}$ (there exist path where a
property eventually holds) and $\mathsf{EX}$ (there exist an immediate
successor satisfying some property).

Furthermore, we extend our construction of the $\pre^*$ images by
showing that the set of predecessors under a regular constraint (i.e.,
the set of all predecessors reachable by computations which stay in
some given regular set of configurations) is also regular and
effectively constructible. For that, we use representation structures
which can be seen as alternating finite-state automata.  This result
allows us to provide a symbolic model checking algorithm for the logic
$\mathsf{E(U,X)}$ with regular atomic predicates, i.e., the fragment
of CTL with the operators $\mathsf{EU}$ (exists until) and
$\mathsf{EX}$ (exists next).

The structure of this paper is the following. In the next two
sections, we introduce higher-order stores and the model of
higher-order context-free processes. We also provide a symbolic
representation for (infinite) \emph{regular} sets of stores using a
certain type of finite automata. Then, for the sake of readability, we
first present our algorithm for computing the unconstrained $\pre$ and
$\pre^*$ sets of a regular set of stores (Section \ref{sect:sra}),
before extending it to the case of $\pre^*$ sets constrained by a
regular set $C$ (Section \ref{sect:csra}). Due to lack of space,
additional definitions and detailed proofs can be found in the full
version of this paper\footnote{available at
  \url{http://www.liafa.jussieu.fr/\~{}ameyer/}.}.

\section{Higher-order Context-free Processes}
\label{sect:hcfp}

We introduce a class of models we call \emph{higher-order context-free
  processes}, which generalize context-free processes (CFP) and are a
subclass of higher-order pushdown systems (HPDS). They manipulate data
structures called \emph{higher-order stores}.

\begin{definition}[Higher-order store]
  The set $\S_1$ of \emph{level $1$ stores} (or \emph{$1$-stores})
  over store alphabet $\Gamma$ is the set of all sequences $[a_1
  \ldots a_l] \in [\Gamma^*]$.  For $n \geq 2$, the set $\S_n$ of
  \emph{level $n$ stores} (or \emph{$n$-stores}) over $\Gamma$ is the
  set of all sequences $[s_1 \ldots s_l] \in [{\S_{n-1}}^+]$.
\end{definition}

\noindent
The following operations are defined on $1$-stores:
\begin{alignat*}{2}
  \push_1^w([a_1 \ldots a_l]) & = [w a_2 \ldots a_l] &\quad& \text{
    for all } w \in \Gamma^*,
  \\
  \top_1([a_1 \ldots a_l]) & = a_1.
  \\
  \intertext{We will sometimes abbreviate $\push_1^\varepsilon$ as
    $\pop_1$. The following operations are defined on $n$-stores ($n >
    1$):} \push_1^w([s_1 \ldots s_l]) & = [\push_1^w(s_1) \ldots s_l]
  \\
  \push_k([s_1 \ldots s_l]) & = [\push_k(s_1) \ldots s_l] & & \text{
    if } k \in [2,n[,
  \\
  \push_n([s_1 \ldots s_l]) & = [s_1 s_1 \ldots s_l]
  \\
  \pop_k([s_1 \ldots s_l]) & = [\pop_k(s_1) \ldots s_l] & & \text{ if
  } k \in [2,n[,
  \\
  \pop_n([s_1 \ldots s_l]) & = [s_2 \ldots s_l] & & \text{ if } l >
  1, \text{ else undefined},
  \\
  \top_k([s_1 \ldots s_l]) & = \top_k(s_1) & & \text{ if } k \in
  [1,n[,
  \\
  \top_n([s_1 \ldots s_l]) & = s_1.
\end{alignat*}
We denote by $O_n$ the set of operations consisting of:
\[
\{\, \push_k, \pop_k\ |\ k \in [2,n] \,\} \cup \{\,\push_1^w\ |\ w \in
\Gamma^*\,\}.
\]
We say that operation $o$ is of level $n$, written $l(o)
= n$, if $o$ is either $\push_n$ or $\pop_n$, or $\push_1^w$ if $n =
1$. We can now define the model studied in this paper.

\begin{definition}
  A \emph{higher-order context-free process} of level $n$ (or
  $n$-HCFP) is a pair $\H = (\Gamma,\Delta)$, where $\Gamma$ is a
  finite alphabet and $\Delta \in \Gamma \times O_n$ is a finite set
  of transitions. A configuration of $\H$ is a $n$-store over
  $\Gamma$. $\H$ defines a transition relation
  $\underset{\H}{\hookrightarrow}$ between $n$-stores (or
  $\hookrightarrow$ when $\H$ is clear from the context), where
  \[
  s \underset{\H}{\hookrightarrow} s' \iff \exists (a,o) \in \Delta
  \text{ such that } \top_1(s) = a \text{ and } s' = o(s).
  \]
\end{definition}

The level $l(d)$ of a transition $d = (a,o)$ is simply the level of
$o$. Let us give a few more notations concerning HCFP computations.
Let $H = (\Gamma,\Delta)$ be a $n$-HCFP.  A \emph{run} of $\H$
starting from some store $s_0$ is a sequence $s_0 s_1 s_2 \ldots$ such
that for all $i \geq 0$, $s_i \hookrightarrow s_{i+1}$. The reflexive
and transitive closure of $\hookrightarrow$ is written
$\overset{*}{\hookrightarrow}$ and called the \emph{reachability}
relation.  For a given set $C$ of $n$-stores, we also define the
\emph{constrained transition} relation $\hookrightarrow_C\ =\ 
\hookrightarrow\ \cap\ (C \times C)$, and its reflexive and
transitive closure $\overset{*}{\hookrightarrow}_C$. Now for any set
of $n$-stores $S$, we consider the sets:
\begin{align*}
  \post_\H[C](S) = \{\,s\ |\ \exists s' \in S,\ s' \hookrightarrow_C s
  \,\},
  \\
  \post^*_\H[C](S) = \{\,s\ |\ \exists s' \in S,\ s'
  \overset{*}{\hookrightarrow}_C s \,\},
  \\
  \pre_\H[C](S) = \{\,s\ |\ \exists s' \in S,\ s \hookrightarrow_C s'
  \,\},
  \\
  \pre^*_\H[C](S) = \{\,s\ |\ \exists s' \in S,\ s
  \overset{*}{\hookrightarrow}_C s' \,\}.
\end{align*}
When $C$ is the set $\S_n$ of all $n$-stores, we omit it in notations
and simply write for instance $\pre_\H(S)$ instead of $\pre_\H[C](S)$.
We will also omit $\H$ when it is clear from the context. When $\H$
consists of a single transition $d$, we may write $\pre_d(S)$ instead
of $\pre_\H(S)$.

\section{Sets of Stores and Symbolic Representation}
\label{sect:symb}

To be able to design symbolic verification techniques over
higher-order context-free processes, we need a way to finitely
represent infinite sets (or languages) of configurations. In this
section we present the sets of configurations (i.e. sets of stores) we
consider, as well as the family of automata which recognize them.

A $n$-store $s = [s_1 \ldots s_l]$ over $\Gamma$ is associated to a
word $w(s) = [w(s_1) \ldots w(s_l)]$, in which store letters in
$\Gamma$ only appear at nesting depth $n$. A set of stores over
$\Gamma$ is called \emph{regular} if its set of associated words is
accepted by a finite automaton over $\Gamma' = \Gamma \cup
\{\,[\,,\,]\,\}$, which in this case we call a \emph{store automaton}.
We will often make no distinction between a store $s$ and its
associated word $w(s)$. Due to the nested structure of pushdown
stores, it will sometimes be more convenient to characterize sets of
stores using \emph{nested store automata}.

\begin{definition}
  A level $1$ nested store automaton is a finite automaton whose
  transitions have labels in $\Gamma$. A nested store automaton of
  level $n \geq 2$ is a finite automaton whose transitions are
  labelled by level $n-1$ nested automata over $\Gamma$.
\end{definition}

\noindent
The existence of a transition labelled by $\B$ between two control
states $p$ and $q$ in a finite automaton $\A$ is written $p
\lra{\A}{\B} q$, or simply $p \lrb{\B} q$ when $\A$ is clear from the
context. Let $\A = (Q,\Gamma,\delta,q_0,q_f)$ be a level $n$ nested
automaton\footnote{ Note that we only consider automata with a single
  final state.} with $n \geq 2$. The level $k$ language of $\A$ for $k
\in [1,n]$ is defined recursively as:
\begin{align*}
  L_k(\A)\ & =\ \{\, [ L_k(\A_1) \ldots L_k(\A_l) ]\ |\ [ \A_1 \ldots
  \A_l ] \in L_n(\A) \,\} & & \text{ if } k < n,
  \\
  L_k(\A)\ & =\ \{\, [ \A_1 \ldots \A_l ]\ |\ q_0\ \lra{\A}{\A_1}
  \ldots \lra{\A}{\A_l} q_f\,\} & & \text{ if } k = n.
\end{align*}
For simplicity, we often abbreviate $L_1(\A)$ as $L(\A)$. We say a
nested automaton $\B$ occurs in $\A$ if $\B$ labels a transition of
$\A$, or occurs in the label of one. Level $n$ automata are well
suited to representing sets of $n$-stores, but have the same
expressive power as standard level $1$ store automata.

\begin{proposition}
  \label{prop:rec}
  The store languages accepted by nested store automata are the
  regular store languages.
\end{proposition}

Moreover, regular $n$-store languages are closed under union,
intersection and complement in $\S_n$. We define for later use the set
of automata $\{\,\A_a^{n}\ |\ a \in \Gamma,\ n \in \mathbb{N}\,\}$
such that for all $a$ and $n$, $L(\A_a^n) = \{\,s \in \S_n\ |\ 
\top_1(s) = a\,\}$. We also write $\A \times \B$ the product operation
over automata such that $L(\A \times \B) = L(\A) \cap L(\B)$.

\section{Symbolic Reachability Analysis}
\label{sect:sra}

Our goal in this section is to investigate effective techniques to
compute the sets $\pre(S)$, $\post(S)$, $\pre^*(S)$ and $\post^*(S)$
for a given $n$-HCFP $\H$, in the case where $S$ is a regular set of
stores. For level $1$ pushdown systems, it is a well-known result that
both $\pre^*_\H(S)$ and $\post^*_\H(S)$ are regular. We will see that
this is still the case for $\pre(S)$ and $\pre^*(S)$ in the
higher-order case, but not for $\post(S)$ (hence not for $\post^*(S)$
either).

\subsection{Forward Reachability}

\begin{proposition}
  \label{prop:post}
  Given a $n$-HCFP $\H$ and a regular set of $n$-stores $S$, the set
  $\post(S)$ is in general not regular. This set is a
  context-sensitive language.
\end{proposition}

\begin{proof}
  Let $\post_{(a,o)}(S)$ denote the set $\{\,s'\ |\ \exists s \in S,\ 
  \top_1(s) = a\ \land\ s' = o(s)\,\}$. Suppose $S$ is a regular set
  of $n$-stores, then if $d = (a,\push_1^w)$ or $d = (a,\pop_k)$, it
  is not difficult to see that $\post_{(a,o)}(S)$ is regular.
  However, if $d = (a,\push_k)$ with $k > 1$, then $\post_{(a,o)}(S)$
  is the set $\{\,[^{n-k+1} t\,t\, w\ |\ [^{n-k+1} t\, w \in S\,\}$.
  It can be shown using the usual pumping arguments that this set is
  not regular, because of the duplication of $t$. However, one can
  straightforwardly build a linearly bounded Turing machine
  recognizing this set. \qed
\end{proof}

\subsection{Backward Reachability}

We first propose a transformation on automata which corresponds to the
$\pre$ operation on their language. In a second time, we extend this
construction to deal with the more difficult computation of $\pre^*$
sets.

\begin{proposition}
  \label{prop:pre}
  Given a $n$-HCFP $\H$ and a regular set of $n$-stores $S$, the
  set $\pre(S)$ is regular and effectively computable.
\end{proposition}

We introduce a construction which, for a given HCFP transition $d$ and
a given regular set of $n$-stores $S$ recognized by a level $n$ nested
automaton $\A$, allows us to compute a nested automaton $\A'_d$
recognizing the set $\pre(S)$ of direct predecessors of $S$ by $d$.
This construction is a transformation over nested automata, which we
call $T_d$. We define $\A'_d = T_d(\A) = (Q', \Gamma, \delta', q_0',
q_f)$ as follows.
\\
If $l(d) < n$, we propagate the transformation to the first level
$n-1$ automaton encountered along each path. We thus have $Q' = Q$,
$q_0' = q_0$ and
\[
\delta' = \{\, q_0 \lrb{T_d(\A_1)} q_1\ |\ q_0 \lra{\A}{\A_1}
q_1\,\} \cup \{\, q \lrb{\B} q'\ |\ q \lra{\A}{\B} q' \land q \neq
q_0\,\}.
\]
If $l(d) = n$, we distinguish three cases according to the nature of
$d$:
\begin{enumerate}
\item If $d = (a,\push_1^w)$, then $Q' = Q \cup \{q'_0\}$ and $\delta'
  = \delta \cup \{\,q'_0 \lrb{a} q_1\ |\ q_0 \lra{\A}{w} q_1\,\}$.
\item If $d = (a,\push_n)$ and $n > 1$, then $Q' = Q \cup \{q'_0\}$
  and
  \\
  $\delta' = \delta \cup \{\,q'_0 \lrb{\B} q_2\ |\ \exists q_1,\ q_0
  \lra{\A}{\A_1} q_1 \lra{\A}{\A_2} q_2\,\}$ where $\B = \A_1 \times
  \A_2 \times \A_a^{(n-1)}$.
\item If $d = (a, \pop_n)$, then $Q' = Q \cup \{q'_0\}$ and $ \delta'
  = \delta \cup \{\,q'_0 \lrb{\A_a^{(n-1)}} q_0\,\}$.
\end{enumerate}
It is not difficult to prove that $L(\A'_d) = \pre_d(L(\A))$. Hence,
if $\Delta$ is the set of transitions of $\H$, then we have $\pre(S) =
\pre(L(\A)) = \bigcup_{d \in \Delta} L(\A'_d)$.

\medskip

This technique can be extended to compute the set $\pre^*(S)$ of all
predecessors of a regular set of stores $S$.

\begin{theorem}
  \label{thm:pre-star}
  Given a $n$-HCFP $\H$ and a regular set of $n$-stores $S$, the
  set $\pre^*(S)$ is regular and effectively computable.
\end{theorem}

To compute $\pre^*(S)$, we have to deal with the problem of
termination. A simple iteration of our previous construction will in
general not terminate, as each step would add control states to the
automaton. As a matter of fact, even the sequence $(\pre^i(S))_{i \geq
  0}$, defined as $\pre^0(S) = S$ and for all $n \geq 1$ $\pre^n(S) =
\pre^{n-1}(S) \cup \pre(\pre^{n-1}(S))$, does not reach a fix-point in
general. For instance, if $d = (a,\pop_1)$, then for all $n$,
$\pre^n([a]) = \{\,[a^i]\ |\ i \leq n\,\} \neq \pre^{n+1}([a])$.

To build $\pre^*(S)$ for some regular $S$, we modify the previous
construction in order to keep constant the number of states in the
nested automaton we manipulate. The idea, instead of creating new
control states, is to add edges to the automaton until saturation,
eventually creating loops to represent at once multiple applications
of a HCFP transition. Then, we prove that this new algorithm
terminates and is correct.

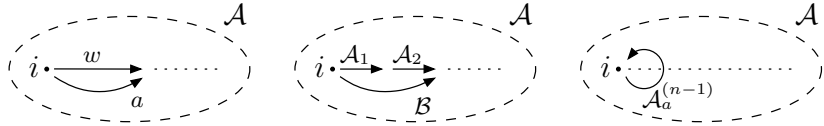
\begin{figure}[htbp]
\begin{center}
  \input{transf.pstex_t}
  \end{center}
  {\caption{transformation $T_d(\A)$ for $d = (a, \push_1^w)$, $(a,
      \push_k)$ and $(a,\pop_k)$.}
    \label{fig:td}}
\end{figure}

Let us first define operation $T_d$ for any $n$-HCFP transition $d$
(see Figure \ref{fig:td} for an illustration). Let $\A = (Q, \Gamma,
\delta, q_0, q_f)$ and $\A' = (Q, \Gamma, \delta', q_0, q_f)$ be
nested $n$-store automata over $\Gamma' = \Gamma \cup \{\,[,]\,\}$,
and $d$ a $n$-HCFP transition.  We define $\A' = T_d(\A)$ as follows.
\\
If the level of $d$ is less than $n$, then we simply propagate the
transformation to the first level $n-1$ automaton encountered along
each path:
\[
\delta' = \{\, q_0 \lrb{T_d(\A_1)} q_1\ |\ q_0 \lra{\A}{\A_1}
q_1\,\} \cup \{\, q \lrb{\B} q'\ |\ q \lra{\A}{\B} q' \land q \neq
q_0\,\}.
\]
If $l(d) = n$ then as previously we distinguish three cases according
to $d$:
\begin{enumerate}
\item If $n = 1$ and $d = (a,\push_1^w)$, then $\delta' = \delta \cup
  \{\,q_0 \lrb{a} q_1\ |\ q_0 \lra{\A}{w} q_1\,\}$.
\item If $d = (a,\push_n)$ for some $n > 1$, then
  \\
  $\delta' = \delta \cup \{\,q_0 \lrb{\B} q_2\ |\ \exists q_1,\ q_0
  \lra{\A}{\A_1} q_1 \lra{\A}{\A_2} q_2\,\}$ where $\B = \A_1 \times
  \A_2 \times \A_a^{(n-1)}$.
\item If $d = (a, \pop_n)$, then $\delta' = \delta \cup \{\,q_0
  \lrb{\A_a^{(n-1)}} q_0\,\}$
\end{enumerate}
Suppose $H = (\Gamma, \Delta)$ with $\Delta = \{\,d_0, \ldots
,d_{l-1}\,\}$. Given an automaton $\A$ such that $S = L(\A)$, consider
the sequence $(\A_i)_{i \geq 0}$ defined as $\A_0 = \A$ and for all $i
\geq 0$ and $j = i \mod l$, $\A_{i+1} = T_{d_j}({\A_i})$. In order to
obtain the result, we have to prove that this sequence always reaches
a fix-point (Lemma \ref{lem:term}) and this fix-point is an automaton
actually recognizing $\pre^*(S)$ (Lemmas \ref{lem:snd} and
\ref{lem:cmpl}).

  

\begin{lemma}[Termination]
  \label{lem:term}
  For all nested $n$-store automaton $\A$ and $n$-HCFP $\H = (\Gamma,
  \Delta)$, the sequence $(\A_i)_{i \geq 0}$ defined with respect to
  $\A$ eventually stabilizes: $\exists k \geq 0,\ \forall k' \in
  \Delta,\ \A_{k'} = \A_k$, which implies $L(\A_k) = \bigcup_{i \geq
    0}L(\A_i)$.
\end{lemma}

\begin{proof}
  First, notice that for all $d$, $T_d$ does not change the set of
  control states of any automaton occurring in $\A$, and only adds
  transitions. This means $(\A_i)_{i \geq 0}$ is monotonous in the
  size of each $\A_i$.
  
  To establish the termination of the conctruction, we prove that the
  number of transitions which can be added to $\A_0$ is finite.  Note
  that by definition of $T_d$, the number of states of each $\A_i$ is
  constant. Moreover, each new transition originates from the initial
  state of the automaton it is added to. Hence, the total number of
  transitions which can be added to a given automaton is equal to
  $|V_n| \cdot |Q|$, where $V_n$ is the level $n$ vocabulary and $Q$
  its set of states. Since $|Q|$ does not change, we only have to
  prove that $V_n$ is finite for all $n$. If $n = 1$, $V_1 = \Gamma$,
  and the property holds. Now suppose $n > 1$ and the property holds
  up to level $n-1$. By induction hypothesis, $V_{n-1}$ is finite.
  With this set of labels, one can build a finite number $N$ of
  different level $n-1$ automata which is exponential in $|V_{n-1}|
  \cdot K$, where $K$ depends on the number of level $n-1$ automata in
  $\A_0$ and of their sets of control states. As each transition of a
  level $n$ automaton is labelled by a product of level $n-1$
  automata, then $|V_n|$ is itself exponential in $N$, and thus doubly
  exponential in $|V_{n-1}|$. Remark that, as a consequence, the
  number of steps of the construction is non-elementary in $n$. \qed
\end{proof}

\begin{lemma}[Soundness]
  \label{lem:snd}
  $\bigcup_{i \geq 0} L(\A_i) \subseteq \pre^*_\H(S).$
\end{lemma}

\begin{proof}[sketch]
  We prove by induction on $i$ the equivalent result that $\forall i,\ 
  L(\A_i) \subseteq \pre^*_\H(S)$. The base case is trivial since by
  definition $\A_0 = \A$ and $L(\A) = S \subseteq \pre^*_\H(S)$. For
  the inductive step, we consider a store $s$ accepted by a run in
  $\A_{i+1}$ and reason by induction on the number $m$ of new level
  $k$ transitions used in this run, where $k$ is the level of the
  operation $d$ such that $\A_{i+1} = T_d(\A_i)$.  The idea is to
  decompose each run containing $m$ new transitions into a first part
  with less than $m$ new transitions, one new transition, and a second
  part also containing less than $m$ new transitions.  Then, by
  induction hypothesis on $m$ and $i$, one can re-compose a path in
  $\A_i$ recognizing some store $s'$ such that $s' \in \pre^*_\H(S)$
  and $s \in \pre^*_\H(s')$. \qed
\end{proof}

\begin{lemma}[Completeness]
  \label{lem:cmpl}
  $\pre^*_\H(S) \subseteq \bigcup_{i \geq 0} L(\A_i).$
\end{lemma}

\begin{proof}[sketch]
  We prove the sufficient property that for all nested store automaton
  $\A$ and HCFP transition $d$, $\pre_{d}(L(\A)) \subseteq
  L(T_d(\A))$. We consider automata $\A$ and $\A'$ such that $\A' =
  T_d(\A)$, and any pair of stores $s \in L(\A)$ and $s' \in
  \pre_{d_j}(s)$. It suffices to isolate a run in $\A$ recognizing $s$
  and enumerate the possible forms of $s'$ with respect to $s$ and $d$
  to be able to exhibit a possible run in $\A'$ accepting $s'$, by
  definition of $T_d$. This establishes the fact that $T_d$ adds to
  the language $L$ of its argument \emph{at least} the set of direct
  predecessors of stores of $L$ by $d$. \qed
\end{proof}

As a direct consequence of Proposition \ref{prop:pre} and Theorem
\ref{thm:pre-star}, we obtain a symbolic model checking algorithm for
the logic $\mathsf{E(F,X)}$ with regular store languages as atomic
predicates, i.e. the fragment of the temporal logic CTL for the modal
operators $\mathsf{EF}$ (there exists a path where eventually a
property holds) and $\mathsf{EX}$ (there exist an immediate successor
satisfying a property).
                                                                                                    
\begin{theorem}
  \label{thm:efx}
  For every HCFP $\H$ and formula $\varphi$ of $\mathsf{E(F,X)}$, the
  set of configurations (stores) satisfying $\varphi$ is regular and
  effectively computable.
\end{theorem}

\section{Constraining Reachability}
\label{sect:csra}

In this section we address the more general problem of computing a
finite automaton recognizing $\pre^*_\H[C](S)$ for any HCFP $H$ and
pair of regular store languages $C$ and $S$. We provide an extension
of the construction of Proposition \ref{thm:pre-star} allowing us to
ensure that we only consider runs of $H$ whose configurations all
belong to $C$. Again, from a given automaton $\A$, we construct a
sequence of automata whose limit recognizes exactly
$\pre^*_\H[C](L(\A))$. The main (and only) difference with the
previous case is that we need to compute language intersections at
each iteration without invalidating our termination arguments (i.e.
without adding any new states to the original automaton). For this
reason, we use a class of \emph{alternating} automata, which we call
\emph{constrained} nested automata.

\begin{definition}[Constrained nested automata]
  Let $\B$ be a non-nested $m$-store automaton\footnote{i.e. a
    standard, level $1$ finite state automaton.} (with $m \geq n$). A
  level $n$ $\B$-constrained nested automaton $\A$ is a nested
  automaton $(Q_\A, \Gamma, \delta_\A, i_\A, f_\A)$ with special
  transitions of the form $p \lra{\A}{\C} (q,r)$ where $p,q \in Q_\A$,
  $r$ is a control state of $\B$ and $\C$ is a level $n-1$
  $\B$-constrained nested automaton.
\end{definition}

For lack of space, we are not able to provide here the complete
semantics of these automata. However, the intuitive idea is quite
simple. Suppose $\A$ is a $\B$-constrained nested $n$-store automaton,
and $\B$ also recognizes $n$-stores.  First, we require all the words
accepted by $\A$ to be also accepted by $\B$: $L(\A) \subseteq L(\B)$.
Then, in any run of $\A$ where a transition of the form $p \lrb{\D}
(q,r)$ occurs, the remaining part of the input word should be accepted
both by $\A$ when resuming from state $q$ and by $\B$ when starting
from state $r$. Of course, when expanding $\D$ into a word of its
language, it may require additional checks in $\B$. As a matter of
fact, constrained nested automata can be transformed into equivalent
level $1$ alternating automata. As such, the languages they accept are
all regular.

\begin{proposition}
  Constrained nested automata accept regular languages.
\end{proposition}

The construction we want to provide needs to refer to whole sets of
paths in a level $1$ store automaton recognizing the constraint
language. To do this, we need to introduce a couple of additional
definitions and notations.

\begin{definition}
  Let $\A$ be a finite store automaton over $\Gamma' = \Gamma \cup
  \{\,[\,,\,]\,\}$. A state $p$ of $\A$ is of level $0$ if it has no
  successor by $[$ and no predecessor by $]$. It is of level $k$ if
  all its successors by $[$ and predecessors by $]$ are of level
  $k-1$. The level of $p$ is written $l(p)$.
\end{definition}

We can show that any automaton recognizing only $n$-stores is
equivalent to an automaton whose control states all have a
well-defined level. A notion of level can also be defined for paths. A
\emph{level $n$ path} in a store automaton is a path $p_1 \ldots p_k$
with $l(p_1) = l(p_k) = n$ and $\forall i \in [2,k-1],\ l(p_i) < n$.
All such paths are labelled by $n$-stores. Now, to concisely refer to
the whole set of level $n$ paths between two level $n$ control states,
we introduce the following notation. Let
\[
Q = \{\,q \in Q_\A\ |\ l(q) < n \land p_1 \lra{\A}{+} q \lra{\A}{+}
p_2\,\}
\]
be the set of all states of $\A$ occurring on a level $n$ path between
$p_1$ and $p_2$. If $Q$ is not empty, we write $p_1 \leads{\A}{\B}
p_2$, where $\B$ is defined as:
\[
\B = \big(\,Q_\B = Q \cup \{p_1,p_2\},\ \Gamma',\ \delta_\B =
\delta_\A \cap (Q_\B \times \Gamma' \times Q_\B),\ p_1,\ p_2\,\big).
\]
Thanks to these few notions, we can state our result:

\begin{theorem}
  \label{thm:pre-starC}
  Given a $n$-HCFP $\H$ and regular sets of $n$-stores $S$ and
  $C$, the set $\pre_\H^*[C](S)$ is regular and effectively
  computable.
\end{theorem}

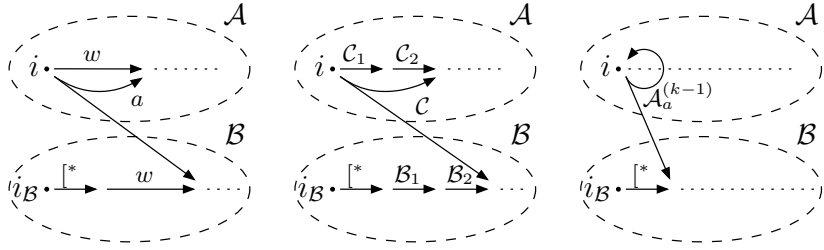
\begin{figure}[htbp]
\begin{center}
  \input{transfB.pstex_t}
  \end{center}
  {\caption{transformation $T_d^\B(\A)$ for $d = (a, \push_1^w)$, $(a,
      \push_k)$ and $(a,\pop_k)$.}
    \label{fig:tdB}}
\end{figure}

To address this problem, we propose a modified version of the
construction of the previous section, which uses constrained nested
automata. Let $d = (a,o)$ be a HCFP transition rule, $\A = (Q_\A,
\Gamma, \delta, i, f)$ and $\A' = (Q_\A, \Gamma, \delta', i, f)$ two
nested $k$-store automata constrained by a level $1$ $n$-store
automaton $\B = (Q_\B, \Gamma', \delta_\B, i_\B, f_\B)$ accepting $C$
(with $n \geq k)$. We define a transformation $T_{d_j}^\B(\A)$, which
is very similar to $T_{d_j}$, except that we need to add alternating
transitions to ensure that no new store is accepted by $\A'$ unless it
is the transformation of a store previously accepted by $\B$ (Cf.
Figure \ref{fig:tdB}). If $l(d) < k$, we propagate the transformation
to the first level $k-1$ automaton along each path:
\[
\delta' = \{\, i \lrb{T_d^\B(\C)} (p,q)\ |\ i \lra{\A}{\C} (p,q)\,\}
\cup \{\, p \lrb{\C} (p',q') \in \delta\ |\ p \neq i\,\}.
\]
If $l(d) = n$, we distinguish three cases according to the nature of
$d$:
\begin{enumerate}
\item If $d = (a,\push_1^w)$, then
  \vspace{-1ex}\begin{multline*}
    \delta' = \delta \cup \big\{\,i \lrb{a} (p,q) \ |\ i
    \lra{\A_i}{w} (p,q')\quad \land \quad \exists q_1,q \in Q_\B,
    \\
    l(q_1) = l(q) = 0,\ i_\B \lra{\B}{[^n} q_1 \lra{\B}{w} q\,\big\}.
  \end{multline*}
\item If $d = (a,\push_k)$, then for $m = n-k+1$ and $\C = (\C_1
  \times \C_2) \times (\B_1 \times \B_2) \times \A_a^{(k-1)}$,
  \vspace{-1ex}\begin{multline*}
    \delta' = \delta \cup \big\{\,i \lrb{\C} (p,q)\ |\ i
    \lra{\A_i}{\C_1} \lra{\A_i}{\C_2} (p,q')\quad \land \quad
    \exists\, q_1,\,q_2,\,q \in Q_\B,
    \\
    l(q_1) = l(q_2) = l(q) = k-1,\ i_\B \lra{\B}{[^m} q_1
    \leads{\B}{\B_1} q_2 \leads{\B}{\B_2} q\,\big\}.
  \end{multline*}
\item If $d = (a,\pop_k)$, then for $m = n-k+1$,
  \vspace{-1ex}
  \[
  \delta' = \delta \cup \big\{\,i \lrb{\A_a^{(k-1)}} (i,q)\ |\ 
  \exists q \in Q_\B,\ l(q) = k-1,\ i_\B \lra{\B_1}{[^m} q\,\big\}.
  \]
\end{enumerate}
Suppose $H = (\Gamma, \Delta)$ with $\Delta = \{\,d_0, \ldots
,d_{l-1}\,\}$. Given an automaton $\A$ such that $S = L(\A)$, consider
the sequence $(\A_i)_{i \geq 0}$ defined as $\A_0 = \A^\B$ (the
$\B$-constrained automaton with the same set of states and transitions
as $\A$, whose language is $L(\A) \cap L(\B)$) and for all $i \geq 0$
and $j = i \mod l$, $\A_{i+1} = T_{d_j}^\B({\A_i})$.  By definition of
$T_d^\B$, the number of states in each $\A_i$ does not vary, and since
the number of control states of $\B$ is finite the same termination
arguments as in Lemma \ref{lem:term} still hold. It is then quite
straightforward to extend the proofs of Lemma \ref{lem:snd} and Lemma
\ref{lem:cmpl} to the constrained case.

This more general construction also allows us to extend Theorem
\ref{thm:efx} to the larger fragment $\mathsf{E(U,X)}$ of CTL, where
formulas can now contain the modal operator $\mathsf{EU}$ (there
exists a path along which a first property continuously holds until a
second property eventually holds) instead of just $\mathsf{EF}$.

\begin{theorem}
  Given a HCFP $\H$ and formula $\varphi$ of $\mathsf{E(U,X)}$, the
  set of configurations (stores) satisfying $\varphi$ is regular and
  effectively computable.
\end{theorem}

\section{Conclusion}
\label{sect-conclu}
                                                                                                    
We have provided an automata-based symbolic technique for backward
reachability analysis of higher-order context-free processes. This
technique can be used to check temporal properties expressed in the
logic $\mathsf{E(U,X)}$. In this respect, our results provide a first
step toward developing symbolic techniques for the model-checking of
higher-order context-free or pushdown processes.
                                                                                                    
Several important questions remain open and are left for future
investigation. In particular, it would be interesting to extend our
approach to the more general case of higher-order pushdown systems,
i.e. by taking into account a set of control states. This does not
seem to be technically trivial, and naive extensions of our
construction lead to procedures which are not guaranteed to terminate.
                                                                                                    
Another interesting issue is to generalize our symbolic approach to
more general properties than reachability and/or safety, including
liveness properties. Finally, it would also be very interesting to
extend our symbolic techniques in order to solve games (such as safety
and parity games) and to compute representations of the sets of all
winning configurations for these games.

\bibliographystyle{plain}


\appendix

\newpage

\section{Appendix}

\subsection{Nested store automata}

\begin{proposition}
  \label{prop:flatten}
  Nested store automata accept regular store languages.
\end{proposition}

\begin{proof}
  We will prove that given a nested store automaton $\A = (Q, \Gamma,
  \delta, i, f)$, one can effectively compute a level $1$ store
  automaton $\flat{\A}$ such that $L_1(\A) = L(\flat{\A})$. We reason
  by induction on the level $n$ of $\A$. For $n=1$, the property
  trivially holds. For greater values of $n$, consider the property as
  true for all levels less than $n$ and let $\A_1 \ldots \A_m$ be the
  level $n-1$ automata labelling the transitions of $\A$. By induction
  hypothesis, we can build level $1$ automata $\flat{\A_1} \ldots
  \flat{\A_m}$ such that $\forall j \in [1,m]$, $L_1(\A_j) =
  L(\flat{\A_j})$. Let $\flat{\A_j} = (Q_j, \Gamma, \delta_j, i_j,
  f_j)$, with all $Q_j$ supposed disjoint. We now build the level $1$
  automaton $\flat{\A} = (Q', \Gamma, \delta', i', f')$ where for all
  $p,q \in Q,\ j \in [1,m],\ r,s,t,u \in Q_j$ and $a \in \Gamma'$ such
  that $p \lra{\A}{\A_j} q$, $i_j \lra{\sflat{\A_j}}{[} r$, $s
  \lra{\sflat{\A_j}}{a} t$ and $u \lra{\sflat{\A_j}}{]} f_j$, we have:
  \begin{align*}
    i' & \lra{\sflat{\A}}{[} i & p & \lra{\sflat{\A}}{[} p r & p s &
    \lra{\sflat{\A}}{a} p t & p u & \lra{\sflat{\A}}{]} q & f &
    \lra{\sflat{\A}}{]} f'
  \end{align*}
  According to this construction, a path of $\flat{\A}$ between two
  control states $p$ and $q$ in $Q \cap Q'$ is labelled by a word $s$
  if and only if $s$ represents a $(n-1)$-store accepted by some
  $\A_j$ such that $p \lra{\A}{\A_j} q$. Hence $\flat{\A}$ accepts all
  words of the form $[s_1 \ldots s_l]$ such that $[\A_{i_1} \ldots
  \A_{i_l}] \in L_n(\A)$ and for all $j$, $s_j \in L(\A_{i_j})$, which
  is precisely the definition of $L(\A)$. \qed
\end{proof}

Before stating the converse, we need to introduce the notion of
\emph{level} of a level $1$ store automaton control state.  Let $\A$
be a finite store automaton over $\Gamma' = \Gamma \cup
\{\,[\,,\,]\,\}$. A state $p$ of $\A$ is of level $0$ if it has no
successor by $[$ and no predecessor by $]$. It is of level $k$ if all
its successors by $[$ and predecessors by $]$ are of level $k-1$. The
level of $p$ is written $l(p)$. We also define a notion of level for
paths. A \emph{level $n$ path} in a store automaton is a path $p_1
\ldots p_k$ with $l(p_1) = l(p_k) = n$ and $\forall i \in [2,k-1],\ 
l(p_i) < n$. All such paths are labelled by $n$-stores. Finally, to
concisely refer to the whole set of level $n$ paths between two level
$n$ control states, we introduce the following notation. Let
\[
Q = \{\,q \in Q_\A\ |\ l(q) < n \land p_1 \lra{\A}{+} q \lra{\A}{+}
p_2\,\}
\]
be the set of all states of $\A$ occurring on a level $n$ path between
$p_1$ and $p_2$. If $Q$ is not empty, we write $p_1 \leads{\A}{\B}
p_2$, where $\B$ is defined as:
\[
\B = \big(\,Q_\B = Q \cup \{p_1,p_2\},\ \Gamma',\ \delta_\B =
\delta_\A \cap (Q_\B \times \Gamma' \times Q_\B),\ p_1,\ p_2\,\big).
\]
Using this notation, we can also very easily translate any level $1$
$n$-store automaton into a level $n$ nested automaton.

\begin{proposition}
  \label{prop:inflate}
  Regular store languages are accepted by nested store automata.
\end{proposition}

\begin{proof}
  Let $\A = (Q, \Gamma', \delta, i, f)$ be a level $1$ automaton
  recognizing $n$-stores. We want to build a level $n$ nested
  automaton $\A' = (Q', \Gamma, \delta', i', f')$ such that $L_1(\A')
  = L(\A)$. As no path of $\A$ labelled by a word which does not
  denote a correct store can be accepting, we may consider without
  loss of generality that the level of every state in $Q$ is
  well-defined. Let $Q_{n-1}$ be the set of level $n-1$ states of
  $\A$. The only states of level $n$ are $i$ and $f$.  If $n = 1$, we
  build $\A'$ with a set of states $Q' = Q_{n-1}$ and the following
  set of transitions:
  \begin{multline*}
    \delta' = \{\,i' \lrb{a} q\ |\ i \lra{\A}{[} p \lra{\A}{a} q\,\}
    \ \cup\ \big(\,\delta\,\cap\, (Q_{n-1} \times \Gamma \times
    Q_{n-1})\big)
    \\
    \ \cup\ \{\,p \lrb{a} f'\ |\ p \lra{\A}{a} q \lra{\A}{]} f\,\}.
  \end{multline*}
  If $n > 1$, for each $p,q \in Q_{n-1}$ and $\B$ such that $p
  \leads{}{\B} q$, we first build inductively a nested automaton $\B'$
  such that $L_1(\B') = L(\B)$. We then give $\A'$ the following set
  of transitions:
  \begin{multline*}
    \delta' = \{\,i' \lrb{\B'} q\ |\ \exists p,q \in Q_{n-1},\ i
    \lra{\A}{[} p \leads{\A}{\B} q\,\}
    \\
    \ \cup\ \{\,p \lrb{\B'} q\ |\ \exists p,q \in Q_{n-1},\ p
    \leads{\A}{\B} q\,\}
    \\
    \ \cup\ \{\,p \lrb{a} f'\ |\ \exists p,q \in Q_{n-1},\ p
    \leads{\A}{a} q \lra{\A}{]} f\,\}.
  \end{multline*}
  A store $s$ is accepted by $\A'$ if and only if there is a path in
  $\A'$ labelled by $\B'_1 \ldots \B'_k$ from $i'$ to $f'$ such that
  $s \in [L_1(\B'_1) \ldots L_1(\B'_k)]$. We thus also have $s \in
  [L(\B_1) \ldots L(\B_k)]$, and hence $s \in L(\A)$. \qed
\end{proof}

\subsection{Reachability.}

We present here more detailed proofs of the soundness and completeness
lemmas for Theorem \ref{thm:pre-star}. 

Before proceeding, we have to present a few additional definitions and
notations. To be able to easily express and manipulate sets of
possible runs of nested automata, we first define the notion of
\emph{store expression}.

\begin{definition}
  A store expression of level $0$ over alphabet $\Gamma$ is simply a
  letter in $\Gamma$. A store expression of level $n > 0$ is either a
  $n$-store $s$, the name $\A$ of a (nested or not) $n$-store
  automaton, a concatenation of level $n$ store expressions, a level
  $n-1$ store expression between square brackets $[e]$, or the
  repeated concatenation $e^+$ of a level $n$ expression $e$.
\end{definition}

Also, to describe runs of nested automata we define a binary relation
$\longmapsto$, which expresses the choice of a particular path in a
nested automaton appearing inside a store expression.

\begin{definition}
  Let $e = u \A v$ be a store expression where $\A$ is a nested
  $n$-store automaton, we write $e\ \longmapsto\ u [ w ] v$ whenever
  $w \in L_n(\A)$. As usual, we write $\maps{*}$ the reflexive and
  transitive closure of $\longmapsto$. A sequence of store expressions
  $e_1 \ldots e_m$ such that $e_1 = \A$, $e_m \in \S_n$ and $\forall i
  \in [1,m-1],\ e_i\ \longmapsto\ e_{i+1}$ is called a run of $\A$.
\end{definition}

Finally, we define a concatenation operation over stores and store
expressions. 

\begin{definition}
  Let $e = [e_1 e_2]$, $f$ and $g$ be store expressions, we write $e =
  f \cdot g$ if either $f = e_1$ and $g = [e_2]$, or $e_1 = f \cdot
  g'$ and $g = [g' e_2]$. Note that if $e$ is a letter in $\Gamma$ or
  an automaton, there are no $f$ and $g$ such that $e = f \cdot g$.
\end{definition}

For instance, we could write $[[a\B][a][bcd]] = a \cdot
[[\B][a][bcd]]$, or $[[a\B][a][bcd]] = [a\B][a] \cdot [[bcd]]$.
Before proving the soundness of the construction of Proposition
\ref{thm:pre-star}, we need a technical lemma expressing the fact
that all cycles on the initial state of a nested automaton during the
computation of $(\A_i)$ correspond to possible runs of the
context-free process we consider.

The following elementary lemma expresses the simple fact that if some
transition $(a,\pop_k)$ can be applied on a certain store, then it
must also be applicable to any store with the same top-most level
$k-1$ store.

\begin{lemma}
  \label{lem:head-off}
  For all HCFP $\H$ and constant\footnote{We say a store expression is
    constant when it contains no automaton.} store expression $s$,
  \[
  \exists t,\ s \cdot t\,\overset{*}{\hookrightarrow}\, t \implies
  \forall t',\ s \cdot t'\, \overset{*}{\hookrightarrow}\, t'.
  \]
\end{lemma}

\begin{proof}
  The proof is a simple induction on the size of expression $s$. \qed
\end{proof}

\begin{lemma}
  \label{lem:snd1} 
  For all $i\geq 0$ and nested $k$-store automaton $\B = (Q, \Gamma,
  \delta, q_0, q_f)$ occurring in $\A_i$, whenever there exist a state
  $q_1 \neq q_0$, path labels $w_1$ and $w_2$, a transition label $\C$
  and a path $q_0 \overset{w_1}{\longrightarrow} q_0
  \overset{\C}{\longrightarrow} q_1 \overset{w_2}{\longrightarrow}
  q_f$ in $\B$, then for all run
  \[
  \A_i \maps{*} \B \cdot r \maps{*} [\,w_1\,\C\,w_2\,] \cdot r
  \maps{*} t \cdot s
  \]
  where $r$ is any store expression, $w_1 \maps{*} t$ and
  $[\,\C\,w_2\,] \cdot r \maps{*} s$, we necessarily have $s \in
  \pre^*_\H(t \cdot s)$ and
  \[
  \A_i \maps{*} \B \cdot r \maps{*} [\,\C\,w_2\,] \cdot r \maps{*} s.
  \]
\end{lemma}

\begin{proof}
  Let us reason by induction on $i$. Assume for simplicity that no
  transition leads to the initial state in any automaton occurring in
  $\A$. If $i=0$, then $w_1 = \varepsilon$ and the property is
  trivial. Now suppose the property is true up to some rank $i \geq
  0$. Call $d$ the level $k$ operation such that $\A_{i+1} =
  T_d(\A_i)$.  Consider the following run $\rho$ of $\A_{i+1}$:
  \[
  \A_{i+1} \maps{*} \B \cdot r \maps{*} [\,w_1\,\C\,w_2\,] \cdot r
  \maps{*} t \cdot s \text{ with } w_1 \maps{*} t.
  \]
  As $w_1$ labels a loop on the initial state of $\B$, another
  possible run of $\A_{i+1}$ is:
  \[
  \A_{i+1} \maps{*} \B \cdot r \maps{*} [\,\C\,w_2\,] \cdot r \maps{*}
  s.
  \]
  We only need to show that $t \cdot s \overset{*}{\hookrightarrow} s$
  to conclude the proof. To do this, we will reason by induction on
  the number $m$ of new level $k$ transitions of $\A_{i+1}$ (i.e.
  transitions of $\A_{i+1}$ not in $\A_i$) used in the $w_1$ cycle on
  $q_0$.
  \begin{list}{}{}
  \item[$m = 0$: ] As $w_1$ contains no new transition, it also labels
    a cycle in $\A_i$. Now, either transition $\C$ belongs to $\A_i$
    or not. In the positive case, $\rho$ is a path in $\A_i$, hence
    the property is true by induction on $i$. In the case where $\C$
    is a new transition, by definition of $\A_{i+1}$, $\A_i$ admits
    the following run:
    \[
    \A_{i} \maps{*} \B \cdot r \maps{*} [\,w_1\,u\,w_2\,] \cdot r
    \maps{*} t \cdot s' \text{ with } w_1 \maps{*} t \text{ and }
    [uw_2] \cdot r \maps{*} s',
    \]
    where $u$ is equal to $\varepsilon$, $\C_1\,\C_2$ or $v$ when $d$
    is $(a,\pop_k)$, $(a,push_k)$ or $(a,push_1^v)$ respectively. By
    induction on $i$, this run verifies the property, hence we have
    \[
    \A_{i} \maps{*} \B \cdot r \maps{*} [\,u\,w_2\,] \cdot r \maps{*}
    s' \text{ with } t \cdot s' \overset{*}{\hookrightarrow} s'.
    \]
    By Lemma \ref{lem:head-off}, this implies that $\forall s'',\ t
    \cdot s'' \overset{*}{\hookrightarrow} s''$, and in particular $t
    \cdot s \hra{\H}{*} s$.
  \item[$m \Rightarrow m+1$:] Suppose the $w_1$ cycle in $\B$ contains
    $m+1$ new transitions. Let $q_0 \lra{\D}{} q_0$ be one of these
    new transitions, we have $w_1 = w'_1 \C w'_2$. Hence $\B$ has a
    path
    \[
    q_0 \lra{\B}{w'_1} q_0 \lra{\B}{\D} q_0 \lra{\B}{w''_1} q_0
    \lra{\B}{\C} q_1 \lra{\B}{w_2} q_f
    \]
    which begins with a cycle on $q_0$ labelled by $w'_1$, containing
    $m$ or less new transitions of $\A_{i+1}$. Suppose $t = t_1 \cdot
    t_2$ and $w'_1 \maps{*} t_1$, by induction hypothesis on $m$ we
    have:
    \[
    \A_{i+1} \maps{*} \B \cdot r \maps{*} [\,\D\,w''_1\,\C\,w_2\,]
    \cdot r \maps{*} t_2 \cdot s
    \]
    and $s \in \pre^*_\H(t_1 \cdot s)$. We now have to examine the way
    transition $\D$ is created in $\A_{i+1}$, which depends on the
    type of $d$. As previously, by definition of $\A_{i+1}$ there must
    be a run of the form
    \[
    \A_{i} \maps{*} \B \cdot r \maps{*} [\,u\,w''_1\,\C\,w_2\,] \cdot
    r \maps{*} t_3 \cdot s,
    \]
    where $u$ is equal to $\varepsilon$, $\D_1\,\D_2$ or $v$ when $d$
    is $(a,\pop_k)$, $(a,push_k^v)$ or $(a,push_1^v)$ respectively. It
    is easy to show that $t_3$ can be chosen to be $d(t_2)$. This run
    uses a path in $\B$ starting with a cycle on $q_0$ labelled by
    $u\,w''_1$ which contains $m$ or less new level $k$ transitions:
    \[
    q_0 \lra{\B}{u} q_0 \lra{\B}{w''_1} q_0 \lra{\B}{\C} q_1
    \lra{\B}{w_2} q_f.
    \]
    Using the induction hypothesis on $m$, we can now conclude that:
    \[
    \A_{i+1} \maps{*} [\,\C\,w_2\,] \cdot r \maps{*} s \text{ and }
    t_3 \cdot s \in \pre^*_\H(s).
    \]
    We have $t \cdot s \hra{\H}{*} t_2 \cdot s \hookrightarrow t_3
    \cdot s \hra{\H}{*} s$, hence $t \cdot s \hra{\H}{*} s$, which
    concludes the proof.\qed
  \end{list}
\end{proof}

\noindent
{\bf Lemma \ref{lem:snd} (Soundness).} {\itshape $\forall i,\ L(\A_i)
  \subseteq \pre^*_\H(S).$}

\begin{proof}
  Assume for simplicity that no transition of an automaton occurring
  in $\A$ leads to its initial state. We reason by induction on $i$.
  The base case is trivial since $\A_0 = \A$ and $L(\A) \subseteq
  \pre^*_\H(L(\A))$. Now consider a store $s$ in $L(\A_{i+1})$. If $s$
  is accepted by $\A_{i+1}$ using no new transition, then it is
  accepted by $\A_i$. Hence by induction hypothesis it belongs to
  $\pre^*_\H(S)$.  Otherwise, the accepting run must be of the form
  \[
  \A_{i+1} \maps{*} \B \cdot r \maps{*} [\,w_1\,\C\,w_2\,] \cdot r
  \maps{*} s,
  \]
  where the path in $\B$ which generates $w_1\,\C\,w_2$ is of the form
  \[
  q_0 \lra{\B}{w_1} q_0 \lra{\B}{\C} q_1 \lra{\B}{w_2} q_f,
  \]
  with $q_1 \neq q_0$. By Lemma \ref{lem:snd1} there exist $t,s_1$
  such that $s = t \cdot s_1$, $t \cdot s_1
  \overset{*}{\hookrightarrow} s_1$ and
  \[
  \A_{i+1} \maps{*} \B \cdot r \maps{*} [\,\C\,w_2\,] \cdot r \maps{*}
  s_1.
  \]
  Note that by definition of $T_d$, all new transitions start from the
  initial states of automata in $\A_{i+1}$. Hence, if the transition
  labelled by $\C$ in the previous run is not new, then the whole run
  exists in $\A_i$. By induction hypothesis on $i$, there exists $s_2
  \in S$ such that $s_1 \overset{*}{\hookrightarrow} s_2$, hence by
  transitivity $s \overset{*}{\hookrightarrow} s_2$.
  \\
  If the transition labelled by $\C$ is new, then since $q_1 \neq q_0$
  and by definition of $T_d$, $d$ must be of the form $(a,\push_k)$ or
  $(a,\push_1^v)$. Then by construction of $\A_{i+1}$ there is a run
  \[
  \A_{i+1} \maps{*} \B \cdot r \maps{*} [\,u\,w_2\,] \cdot r \maps{*}
  s_2,
  \]
  where $u$ is either $\C_1\,\C_2$ if $k > 1$ or $v$ is $k = 1$, and
  $s_2$ can be chosen as $d(s_1)$. Now by induction hypothesis on $i$,
  there exists $s_3 \in S$ such that $s_2 \overset{*}{\hookrightarrow}
  s_3$, hence by transitivity $s \overset{*}{\hookrightarrow} s_3$.
  \qed
\end{proof}

\noindent
{\bf Lemma \ref{lem:cmpl} (Completeness).}  {\itshape For all nested
  store automaton $\A$ and HCFP transition $d$, $\pre_{d}(L(\A)
  \subseteq L(T_d(\A))$.}

\begin{proof}
  Let $\A' = T_d(\A)$. Consider a store $s \in L(\A)$, and let $s'$ be
  any store such that $s' \in \pre_{d_j}(s)$. There is a run $\rho$ of
  $\A$ recognizing $s$ as follows:
  \[
  \A \maps{*} \B \cdot r \longmapsto [\,\C_1 \ldots \C_l] \cdot r
  \maps{*} s.
  \]
  Depending on $d$, we have to consider three cases:
  \begin{enumerate}
  \item If $d_j = (a,\pop_k)$, then $s' = t \cdot s$ where $t$ is any
    store of level $k-1$ such that $\top_1(t) = a$, and by definition
    of $T_d$ the following run exists:
    \[
    \A' \maps{*} [\,\A_a^{(k-1)} \C_1 \ldots \C_l\,] \cdot r \maps{*}
    s'.
    \]
  \item If $d_j = (a,\push_k),\ k>0$, then $s = t t \cdot r$ and $s' =
    t \cdot r$ where $\top_1(t) = a$ and $t$ is in both $L(\C_1)$ and
    $L(\C_2)$. Hence $t$ is also accepted by the level $k-1$ automaton
    $\C_1 \times \C_2 \times \A_a^{k-1}$. Thus, by definition of $T_d$ the
    following run exists:
    \[
    \A' \maps{*} [\,\C_1 \times \C_2 \times \A_a^{k-1}\, \C_3 \ldots
    \C_l\,] \cdot r \maps{*} s'.
    \]
  \item If $d_j = (a,\push_1^w)$, then $s = w \cdot r$ and $s' = a
    \cdot r$. This means $\C_1 \ldots \C_l$ are level $0$ automata
    (i.e. letters), and $\C_1 \ldots \C_{|w|} = w$. By definition of
    $T_d$ the following run exists:
    \[
    \A' \maps{*} [\,a\, \C_{|w|+1} \ldots \C_l\,] \cdot r \maps{*} s'.
    \]
  \end{enumerate}
  This establishes the fact that $T_d$ adds to the language $L$ of its
  argument \emph{at least} the set of direct predecessors of stores of
  $L$ by operation $d$. \qed
\end{proof}

\subsection{Constrained nested automata.}

The language of a constrained nested automaton is defined \emph{via} a
simple adaptation of the construction of Prop. \ref{prop:flatten}.
Consider a nested automaton $\A = (Q, \Gamma, \delta, i, f)$ of level
$n$ constrained with respect to a level $1$ $n$-store automaton $\B =
(Q_\B, \Gamma', \delta_\B, i_\B, f_\B)$ \footnote{note that the levels
  of $\A$ and $\B$ have to be the same for $L(\A)$ to be defined.}.
First, consider the (unconstrained) nested automaton $\A' = (Q,
\Gamma, \delta', i, f)$, where $\delta' = \{\,p \lrb{\C} q\ |\ p
\lra{\A}{\C} (q,r)\,\}$. Second, build according to the construction
of Prop. \ref{prop:flatten} a level $1$ automaton $\flat{\A'} =
(\flat{Q'}, \Gamma', \flat{\delta'}, i', f')$ with the same accepted
language as $\A'$. By adding to $\flat{\A'}$ the control states of
$\B$ and integrating into it the set of constrained transitions of
$\A$, one gets an alternating store automaton $\flat{\A} = (\flat{Q},
\Gamma', \flat{\delta}, (i' \land i_\B), f')$, where $\flat{Q} =
\flat{Q'} \cup Q_\B$. By construction, control states in $\flat{Q'}$
are of the form $q_n \ldots q_k$ where $k \in [1,n]$ and each $q_i$ is
a control state of a level $k$ automaton occurring in $\A'$. We define
$\flat{\delta}$ as the union of $\delta_\B$ and the set of all $s
\lrb{x} t$ such that:
\begin{enumerate}
\item $\bar{p} p r \lrb{x} \bar{p} q \in \flat{\delta'}$, $s =
  \bar{p} p r$, $t = (\bar{p} q \land u)$, $X = ]$ and $p \lra{\C}{\D}
  (q,u)$ where $\C$ occurs in $\A$ and $r$ is a control state of $\D$,
\item $\bar{p} p \lrb{x} \bar{p} q' \in \flat{\delta'}$, $s =
  \bar{p} p$, $t = (\bar{p} q \land u)$, $X = a$, and $p \lra{\C}{a}
  (q,u)$ where $\C$ is a level $1$ automaton occurring in $\A$,
\item $s \lrb{x} t \in \flat{\delta'}$ in all other cases.
\end{enumerate}
We now define the language accepted by $\A$ as the language accepted
by the alternating automaton $\flat{\A}$ we just defined, according to
the usual notion of acceptance for alternating automata: $L(\A) =
L(\flat{\A})$ (please note that the initial state of $\flat{\A}$ is
$i' \land i_\B$).

\subsection{Constrained reachability.}

We give here three lemmas allowing to prove the correctness of the
construction in Section \ref{sect:csra}.

\begin{lemma}[Termination]
  For all nested $n$-store automaton $\A$, level $1$ $n$-store
  automaton $\B$ and $n$-HCFP $\H = (\Gamma, \Delta)$, the sequence
  $(\A^\B_i)$ defined with respect to $\A$ and $\B$ eventually reaches
  $T^\B_\H(\A)$:
  \[ 
  \exists k \geq 0,\ \forall d \in \Delta,\ T^\B_d(\A^\B_k) = \A^\B_k.
  \]  
\end{lemma}

\begin{proof}
  The algorithm for computing $T^\B_\H(\A)$ is similar to the one for
  computing $T_\H(\A)$, except that it labels some of the transitions
  of each $\A^\B_i$ by a state of $\B$. As the number of such states
  remains unchanged throughout the whole computation, this does not
  add any unboundedness in the computation and the maximal number of
  iterations before reaching a fix-point is still finite.  \qed
\end{proof}

\begin{lemma}[Soundness]
  $\forall i,\ L(\A^\B_i) \subseteq \pre^*_\H[C](S).$
\end{lemma}

\begin{proof}
  By definition of $L(\A^\B_i)$, $L(\A^\B_i) \subseteq L(\A_i)$ for
  all $i$. So, by Lemma \ref{lem:snd}, we already have $L(\A^\B_i)
  \subseteq \pre^*_\H(S)$. Let us reason by induction on $i$. By
  definition of constrained nested automata, $L(\A^\B_0) = L(\A) \cap
  C$, hence $L(\A^\B_0) \subseteq \pre^*_\H[C](S)$. Now assume the
  property is true up to some rank $i$, and consider the automaton
  $\A^\B_{i+1}$. Note that everywhere transformation $T^\B_d$ adds a
  transition in $\A^\B_i$ to get $\A^\B_{i+1}$, the alternating
  transitions induced in $\flat{\A^\B_{i+1}}$ ensure that each store
  labelling a new accepting path in the automaton is a transformation
  of a store labelling an accepting path in $\B$. This way, one makes
  sure that no element of $C$ in $\pre^*_\H(S) \setminus
  \pre^*_\H[C](S)$ is added to the language of $\A^\B_{i+1}$.
  
  For instance, assume $d = (a,\push_1^w)$ and some store $s$ is
  accepted by $\flat{\A^\B_{i+1}}$ using a $a$-transition newly
  created by $T_d^\B$. According to the definition of $T_d^\B$, this
  transition is of the form $p \lrb{a} (q,r)$, where $r$ is a
  control state reachable in $\B$ through a path labelled by $[^n\,w$.
  Thus, if we let $w(s) = [^n\,a\,w'$, for $s$ to be accepted by
  $\flat{\A^\B_{i+1}}$, then necessarily $s'$ must be accepted by $\B$
  from state $r$. The same kind of reasoning holds for the other types
  of operations. \qed
\end{proof}

\begin{lemma}[Completeness]
  $\forall i,\ S^C_i \subseteq L(\A^\B_{i}).$
\end{lemma}

\begin{proof}
  By definition, $L(\A^\B_{0}) = S \cap C = S^C_0$. Now suppose the
  property is true up to some rank $i$, and consider a store $s \in
  S^C_{i+1} \setminus S^C_i$. Let $d$ be the operation such that
  $S^C_{i+1} = S^C_i \cup (d(S^C_i) \cap C)$. By definition, there is
  a store $s' \in S^C_i$ such that $s' = d(s)$, and by induction
  hypothesis $s'$ is accepted by $\A^\B_{i}$. Moreover, since both $s$
  and $s'$ are in $C$, they are accepted by $\B$. As seen in Lemma
  \ref{lem:cmpl}, transformation $T_d$ adds a new transition $p_0
  \lrb{\C} q$ creating in particular a path labelled by $s$. The
  additional constraints $T_d^\B$ puts on this transition, and all
  paths in $\A^\B_{i+1}$ in general, forbids any path labelled by some
  $r$ using this transition to be accepted unless both $r$ and $d(r)$
  also have an accepting run in $\B$. This is the case for $s$ and
  $s'$, hence $s \in L(\A^\B_{i+1})$.\qed
\end{proof}

\end{document}

%% file: transf.pstex_t
\begin{picture}(0,0)%
\includegraphics{transf.pstex}%
\end{picture}%
\setlength{\unitlength}{4144sp}%
\begingroup\makeatletter\ifx\SetFigFont\undefined%
\gdef\SetFigFont#1#2#3#4#5{%
  \reset@font\fontsize{#1}{#2pt}%
  \fontfamily{#3}\fontseries{#4}\fontshape{#5}%
  \selectfont}%
\fi\endgroup%
\begin{picture}(4876,742)(353,-1736)
\put(2251,-1456){\makebox(0,0)[rb]{\smash{{\SetFigFont{12}{14.4}{\familydefault}{\mddefault}{\updefault}{\color[rgb]{0,0,0}$i$}%
}}}}
\put(3421,-1141){\makebox(0,0)[b]{\smash{{\SetFigFont{11}{13.2}{\familydefault}{\mddefault}{\updefault}{\color[rgb]{0,0,0}$\A$}%
}}}}
\put(541,-1456){\makebox(0,0)[rb]{\smash{{\SetFigFont{12}{14.4}{\familydefault}{\mddefault}{\updefault}{\color[rgb]{0,0,0}$i$}%
}}}}
\put(1711,-1141){\makebox(0,0)[b]{\smash{{\SetFigFont{11}{13.2}{\familydefault}{\mddefault}{\updefault}{\color[rgb]{0,0,0}$\A$}%
}}}}
\put(1126,-1636){\makebox(0,0)[b]{\smash{{\SetFigFont{9}{10.8}{\familydefault}{\mddefault}{\updefault}{\color[rgb]{0,0,0}$a$}%
}}}}
\put(856,-1366){\makebox(0,0)[b]{\smash{{\SetFigFont{9}{10.8}{\familydefault}{\mddefault}{\updefault}{\color[rgb]{0,0,0}$w$}%
}}}}
\put(3961,-1456){\makebox(0,0)[rb]{\smash{{\SetFigFont{12}{14.4}{\familydefault}{\mddefault}{\updefault}{\color[rgb]{0,0,0}$i$}%
}}}}
\put(5131,-1141){\makebox(0,0)[b]{\smash{{\SetFigFont{11}{13.2}{\familydefault}{\mddefault}{\updefault}{\color[rgb]{0,0,0}$\A$}%
}}}}
\put(2431,-1366){\makebox(0,0)[b]{\smash{{\SetFigFont{9}{10.8}{\familydefault}{\mddefault}{\updefault}{\color[rgb]{0,0,0}$\A_1$}%
}}}}
\put(2746,-1366){\makebox(0,0)[b]{\smash{{\SetFigFont{9}{10.8}{\familydefault}{\mddefault}{\updefault}{\color[rgb]{0,0,0}$\A_2$}%
}}}}
\put(2836,-1681){\makebox(0,0)[b]{\smash{{\SetFigFont{9}{10.8}{\familydefault}{\mddefault}{\updefault}{\color[rgb]{0,0,0}$\B$}%
}}}}
\put(4366,-1636){\makebox(0,0)[b]{\smash{{\SetFigFont{9}{10.8}{\familydefault}{\mddefault}{\updefault}{\color[rgb]{0,0,0}$\A_a^{(n-1)}$}%
}}}}
\end{picture}%

%% file: transfB.pstex_t
\begin{picture}(0,0)%
\includegraphics{transfB.pstex}%
\end{picture}%
\setlength{\unitlength}{4144sp}%
\begingroup\makeatletter\ifx\SetFigFont\undefined%
\gdef\SetFigFont#1#2#3#4#5{%
  \reset@font\fontsize{#1}{#2pt}%
  \fontfamily{#3}\fontseries{#4}\fontshape{#5}%
  \selectfont}%
\fi\endgroup%
\begin{picture}(4876,1460)(353,-2454)
\put(2251,-2176){\makebox(0,0)[rb]{\smash{{\SetFigFont{12}{14.4}{\familydefault}{\mddefault}{\updefault}{\color[rgb]{0,0,0}$i_\B$}%
}}}}
\put(2431,-1366){\makebox(0,0)[b]{\smash{{\SetFigFont{9}{10.8}{\familydefault}{\mddefault}{\updefault}{\color[rgb]{0,0,0}$\C_1$}%
}}}}
\put(2746,-1366){\makebox(0,0)[b]{\smash{{\SetFigFont{9}{10.8}{\familydefault}{\mddefault}{\updefault}{\color[rgb]{0,0,0}$\C_2$}%
}}}}
\put(2251,-1456){\makebox(0,0)[rb]{\smash{{\SetFigFont{12}{14.4}{\familydefault}{\mddefault}{\updefault}{\color[rgb]{0,0,0}$i$}%
}}}}
\put(3421,-1141){\makebox(0,0)[b]{\smash{{\SetFigFont{11}{13.2}{\familydefault}{\mddefault}{\updefault}{\color[rgb]{0,0,0}$\A$}%
}}}}
\put(3421,-1861){\makebox(0,0)[b]{\smash{{\SetFigFont{11}{13.2}{\familydefault}{\mddefault}{\updefault}{\color[rgb]{0,0,0}$\B$}%
}}}}
\put(2431,-2086){\makebox(0,0)[b]{\smash{{\SetFigFont{9}{10.8}{\familydefault}{\mddefault}{\updefault}{\color[rgb]{0,0,0}$[^*$}%
}}}}
\put(2746,-2086){\makebox(0,0)[b]{\smash{{\SetFigFont{9}{10.8}{\familydefault}{\mddefault}{\updefault}{\color[rgb]{0,0,0}$\B_1$}%
}}}}
\put(3061,-2086){\makebox(0,0)[b]{\smash{{\SetFigFont{9}{10.8}{\familydefault}{\mddefault}{\updefault}{\color[rgb]{0,0,0}$\B_2$}%
}}}}
\put(2836,-1681){\makebox(0,0)[b]{\smash{{\SetFigFont{9}{10.8}{\familydefault}{\mddefault}{\updefault}{\color[rgb]{0,0,0}$\C$}%
}}}}
\put(541,-1456){\makebox(0,0)[rb]{\smash{{\SetFigFont{12}{14.4}{\familydefault}{\mddefault}{\updefault}{\color[rgb]{0,0,0}$i$}%
}}}}
\put(1711,-1141){\makebox(0,0)[b]{\smash{{\SetFigFont{11}{13.2}{\familydefault}{\mddefault}{\updefault}{\color[rgb]{0,0,0}$\A$}%
}}}}
\put(1711,-1861){\makebox(0,0)[b]{\smash{{\SetFigFont{11}{13.2}{\familydefault}{\mddefault}{\updefault}{\color[rgb]{0,0,0}$\B$}%
}}}}
\put(721,-2086){\makebox(0,0)[b]{\smash{{\SetFigFont{9}{10.8}{\familydefault}{\mddefault}{\updefault}{\color[rgb]{0,0,0}$[^*$}%
}}}}
\put(1126,-1636){\makebox(0,0)[b]{\smash{{\SetFigFont{9}{10.8}{\familydefault}{\mddefault}{\updefault}{\color[rgb]{0,0,0}$a$}%
}}}}
\put(856,-1366){\makebox(0,0)[b]{\smash{{\SetFigFont{9}{10.8}{\familydefault}{\mddefault}{\updefault}{\color[rgb]{0,0,0}$w$}%
}}}}
\put(1171,-2086){\makebox(0,0)[b]{\smash{{\SetFigFont{9}{10.8}{\familydefault}{\mddefault}{\updefault}{\color[rgb]{0,0,0}$w$}%
}}}}
\put(541,-2176){\makebox(0,0)[rb]{\smash{{\SetFigFont{12}{14.4}{\familydefault}{\mddefault}{\updefault}{\color[rgb]{0,0,0}$i_\B$}%
}}}}
\put(3961,-2176){\makebox(0,0)[rb]{\smash{{\SetFigFont{12}{14.4}{\familydefault}{\mddefault}{\updefault}{\color[rgb]{0,0,0}$i_\B$}%
}}}}
\put(3961,-1456){\makebox(0,0)[rb]{\smash{{\SetFigFont{12}{14.4}{\familydefault}{\mddefault}{\updefault}{\color[rgb]{0,0,0}$i$}%
}}}}
\put(5131,-1141){\makebox(0,0)[b]{\smash{{\SetFigFont{11}{13.2}{\familydefault}{\mddefault}{\updefault}{\color[rgb]{0,0,0}$\A$}%
}}}}
\put(4366,-1636){\makebox(0,0)[b]{\smash{{\SetFigFont{9}{10.8}{\familydefault}{\mddefault}{\updefault}{\color[rgb]{0,0,0}$\A_a^{(k-1)}$}%
}}}}
\put(5131,-1861){\makebox(0,0)[b]{\smash{{\SetFigFont{11}{13.2}{\familydefault}{\mddefault}{\updefault}{\color[rgb]{0,0,0}$\B$}%
}}}}
\put(4141,-2086){\makebox(0,0)[b]{\smash{{\SetFigFont{9}{10.8}{\familydefault}{\mddefault}{\updefault}{\color[rgb]{0,0,0}$[^*$}%
}}}}
\end{picture}%